**Modelling the Relationship Between Post Encroachment Time and Signal Timings Using UAV Video data**

**Zubayer Islam, Ph.D.**
Post-Doctoral Scholar
Department of Civil, Environmental and Construction Engineering
University of Central Florida, Orlando, FL 32816, USA
Email: zubayer_islam@knights.ucf.edu

**Mohamed Abdel-Aty Ph.D., P.E.**
Pegasus Professor and Chair
Department of Civil, Environmental and Construction Engineering
University of Central Florida, Orlando, FL 32816, USA
Tel: (407)823-4535
Email: M.Aty@ucf.edu

**Amrita Goswamy, Ph.D.**
Post-Doctoral Scholar
Department of Civil, Environmental and Construction Engineering
University of Central Florida, Orlando, FL 32816, USA
Email: amrita.goswamy@ucf.edu

**Amr Abdelraouf, Ph.D.**
Post-Doctoral Scholar
Department of Civil, Environmental and Construction Engineering
University of Central Florida, Orlando, FL 32816, USA
Email: amr.abdelraouf@Knights.ucf.edu

**Ou Zheng**
Department of Civil, Environmental and Construction Engineering
University of Central Florida, Orlando, FL 32816, USA
Email: ouzheng1993@Knights.ucf.edu

## ABSTRACT

Intersection safety often relies on the correct modelling of signal phasing and timing parameters. A slight increase in yellow time or red time can have significant impact on the rear end crashes or conflicts. This paper aims to identify the relationship between surrogate safety measures and signal phasing. Unmanned Aerial Vehicle (UAV) video data has been used to study an intersection. Post Encroachment Time (PET) between vehicles was calculated from the video data as well as speed, heading and relevant signal timing parameters such as all red time, red clearance time, yellow time, etc. Random Parameter Ordered Logit Model was used to model the relationship between PET and these signal timing parameters. Overall, the results showed that yellow time and red clearance time is positively related to PETs. The model was also able to idendity certain signal phases that could be a potential safety hazard and would need to be retimed by considering the PETs. The odds ratios from the models also indicates that increasing the yellow and red clearance times by one second can improve the PET levels by 16% and 3% respectively.

**INTRODUCTION**

Traffic analysis from a safety point of view has largely relied on crash data. Various statistical methods and machine learning methods have been implemented to understand proactive natures of crash enabling real time prediction of these events. Countermeasures have been developed based on accident data as well. However, crash data can be rare events and thereare notable shortcomings of these types of data such as incorrect reasoning, subjectivism, inaccurate data, etc. (1, 2). Moreover, the specific reasoning to a crash can often be factors other than roadway characteristics and traffic features which cannot be modelled using the prediction algorithms in the literature. On the other hand, conflict events are more common and therefore, can help better to understand design flaws of roadway as well as traffic conditions that impacts conflicts. Several previous studies have definitively proven conflict analysis as an alternative to crash analysis with similar results  (2-5). Several metrics has thus been developed to measure conflict such as Time-to-collision (TTC) (6), time exposed TTC (TET), time integrated TTC (TIT), time-to-lane crossing (TLC) (7), Post encroach time (PET), gap time (GT), encoding time (ET), and time advantage (TAdv) (8), etc.

The surrogate safety measures are usually dependent on exact localization of road users. For example, to calculate TTC, initial location and velocity would be needed. This requires precise GPS locations. An effective way to study an intersection would be with the help of an Unmanned Aerial Vehicle (UAV) that can be then used to extract accurate trajectories at the centimeter level. These are a better alternative than roadside cameras which have distortion of localization at camera edges. UAVs are also known for easy maneuvering, flexibility, and low cost. UAVs have become an emerging video analysis solution at the transportation level in the recent years. It is often augmented with radar and infrared cameras that can provide a bird's eye view of an intersection including the approaches. In this study, an intersection was analyed with respect to PET from the data available through UAV. The signal timing at that instant was also captured. The purpose of this study was to analyze the interaction of safety events and relate it to the signal states.

**LITERATURE REVIEW**

Traffic safety at intersections has been shown to be dependent on signal timing at that intersection. For example, altering signal phases can better or worsen intersection safety (9). Several studies have found that there is a direct relation between signal timings and crashes. After any retiming of signals, a crash reduction factor is also estimated but few studies have also reported that there were no significant relationships (10). Guo, Wang (11) showed that adaptive intersections experienced fewer crashes than isolate ones. The study was extensive and included over 170 intersections in Florida, USA but the results were based on signal timing sheets only since real traffic data was not available. Midenet, Saunier (12) evaluated signal safety by measuring the exposure to lateral collisions using video feed. Approach level data from traffic detectors including speed, volume was found to be associated with significant crash risk (13). It was also reported in this study that longer green time for left turn, higher green ratio can improve the safety at intersections. The main limitation of all the studies is that crash events are usually rare and therefore, these studies would only rely on the spatial relationship between crash events and traffic parameters. It has been shown in several studies that the temporal relationship need to be included as well since traffic parameters and signal timing would vary largely throughout the day and even across days (14-16). Moreover, there are notable shortcomings of these types of police reported crash data such as incorrect reasoning, subjectivism, inaccurate data, etc. (1, 2). Additionally, there is the moral dilemma of waiting for fatalities to happen before taking an appropriate countermeasure making it a reactive approach. Crash events are also rare, and it takes a long time to study a location or conduct a before-after study. Surrogate safety measures provide an alternate and proactive methodology that does not require much time and solves the moral dilemma to a great extent. Several studies have also shown that it can significantly correlate to crashes and can mostly be used as an alternative (2-5, 17).

Using surrogate safety measures for signal timing was first proposed by Stevanovic, Stevanovic (18). The study proposed the integration of optimization and surrogate safety measure assessment at the microscopic level considering both the safety and efficiency. Network wide optimization was also studied in recent time (19). This work also incorporates simulation and surrogate safety measures to find optimal solution using a model callibrated from real-world data. The influence of signal phasing on the safety and traffic smoothness was also stuided (20, 21). It was also shown that optimization of the left turn waiting zones would improve capacity without degrading traffic flow (22) while Lin and Huang (23) improved both at signal coordination level across multiple intersections. All the studies have relied on simulation softwares such as VISSIM to model traffic signals and safety. While some studies calibrate the models based on real traffic flow, the ground data can be siginifically different than the simulation. This work addresses this research gap and uses real-world data from UAVs to evaluate signal timing based on Post Encroachment Time (PET). The main objective of this work was to evaluate the impact of all-red time, red clearance time, red time, yellow time and green time on the surrogate safety measures based on real-world data. These can also help relevant authorities to understand intersection traffic with respect to PET and gain insight whether the signal timing need optimization or not. Moreover, the odds ratio was also calculated to show that one second increase of yellow and red clearance time will help to increase the PET level thereby improving the safety condition of the intersection.

## DATA PREPARATION
### Trajectory Data

The vehicle trajectories provided by the CitySim dataset (24) were utilized to identify, process, and analyze PET conflicts in this study. The CitySim dataset is composed of top-view drone-video-based vehicle trajectories. The authors identified vehicle trajectories using mask-RCNN and subsequently extracted and exported rotation-aware bounding boxes. The dataset contains vehicle trajectories sampled at 30 frames per second. For each trajectory point, the dataset provides four bounding box positions, speed, and heading. In this work, the University@Alafaya intersection location was selected for development, evaluation, and analysis. The intersection geometry is illustrated in **Figure 1**. It is a signalized intersection between Alafaya Trail (9 lanes) and University Boulevard (9 lanes). The utilized trajectories were extracted from a video recorded on a weekday between 5:40 PM and 6:40 PM (afternoon peak). A total of 4871 vehicles passed through the intersection during that period of time. The different phases are also shown in Figure 1. There are three through lanes for each of the phases 2,4,6 and 8 while two left turning lanes for phases 1,3,5 and 7. The approach 4 does not have any exclusice right turn lanes while the other through phases all have an exclusive right turn lane.

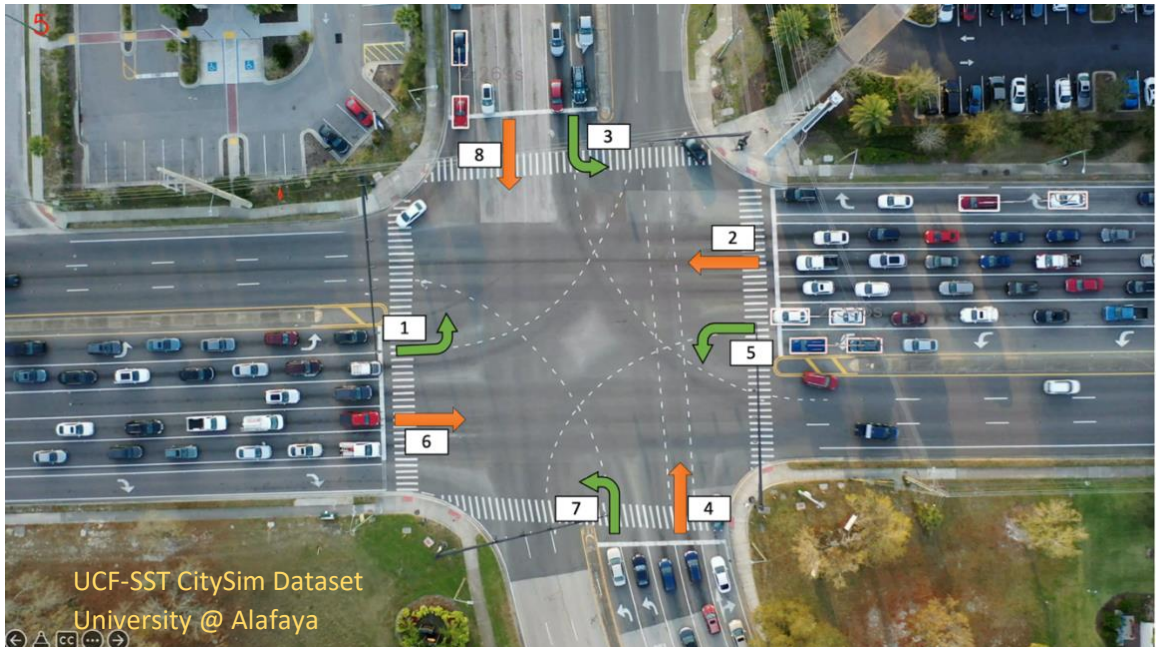

**Figure 1. Study intersection location showing the different phases**

**Post Encroachment Time (PET)**

Post Encroachment Time is a conflict indicator that serves as a surrogate safety measure.

**Figure 2** depicts an example PET conflict between two vehicles at a single timestep. PET measures the period of time between a leading vehicle leaving a particular location and a lagging vehicle arriving at the same location. In this scenario, the location where both vehicles interact is dubbed the conflict zone. The PET conflict indictor generates a sequence of PET values that describe the serial interaction between two vehicle trajectories under observation. A PET value exists in the generated PET sequence as long as the lagging vehicle remains in a conflict zone. Otherwise, the PET value at a timestep where no encroachment occurs is undefined.

In this research effort, the PET values were computed using the rotation-aware vehicle bounding boxes provided by the CitySim Dataset. At each timestep, and for each possible pair of vehicles, the PET value was measured between the moment a lagging vehicle bounding box intersects with a leading vehicle's previous bounding box location (i.e., the lagging vehicle intersects with the conflict zone as described in

**Figure 2**). For each pair of vehicles, an output PET sequence that describes their interaction was generated. The selected timestep was 1/3 seconds (3Hz). PET values under 5 seconds were recorded.

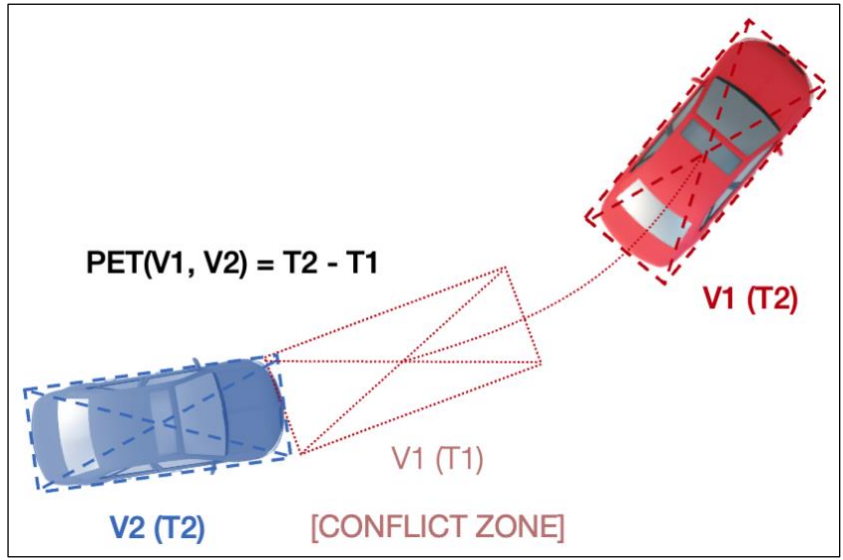

**Figure 2. Example PET calculation between leading vehicle (V1) and lagging vehicle (V2) in the time period between (T1) – (T2) in the conflict zone V1(T1)**

**Table 1** describes the PET conflicts extracted from the study area. When sampled at 3Hz, a total of 193,000 PET conflicts under 5 seconds were captured in the study area. Additionally, **Table 1** reports the minimum PETs (minPETs). The minPET is defined as the minimum PET recorded between 2 vehicle trajectories. It describes the single most hazardous moment between unique vehicle pairs. **Table 1** indicates that, during the recorded time, 717 unique vehicle pairs recorded a minPET under 1 second, and 7345 unique vehicle pairs encountered a minPET conflict under 5 seconds.

**Table 1. Number of PET conflicts and minPET values for different thresholds**

| PET Threshold | < 1.0s | < 2.0s | < 3.0s | < 4.0s | < 5.0s |
|---|---|---|---|---|---|
| Number of PET conflicts (sampled at 3 Hz) | 9K | 62K | 106K | 150K | 193K |

| Number of minPETs for unique vehicle pairs | 717 | 2785 | 4365 | 5897 | 7345 |
|---|---|---|---|---|---|

Utilizing the vehicle bounding boxes for PET calculation is not common within previous research efforts. Instead, most previous work relied on center-point-based conflict identification. As illustrated in **Figure 2**, the vehicle geometry is essential for robust PET measurement. Center points misrepresent vehicle geometries and lead the conflict identification algorithm to neglect conflicts or underestimate their severity (25). **Figure 3** compares heatmap plots of minPETs recorded in the study intersection using bounding boxes versus center points. It can be clearly observed that the bounding box approach was able to recall more conflicts than the center point method. For a minPET < 1.0 second, the center point method identified 141 compared to 717 conflicts captured by the bounding box. Similarly, for a minPET maximum threshold of 3.0 seconds, the center point and bounding box methods identified 3637 and 4365 conflicts, respectively. **Figure 3** clearly demonstrates the superiority and robustness of the bounding box approach. Furthermore, it indicates that the center point misdetection rate is proportional to the conflict severity, meaning that center-point-based computations fail to capture the most hazardous traffic conflicts.

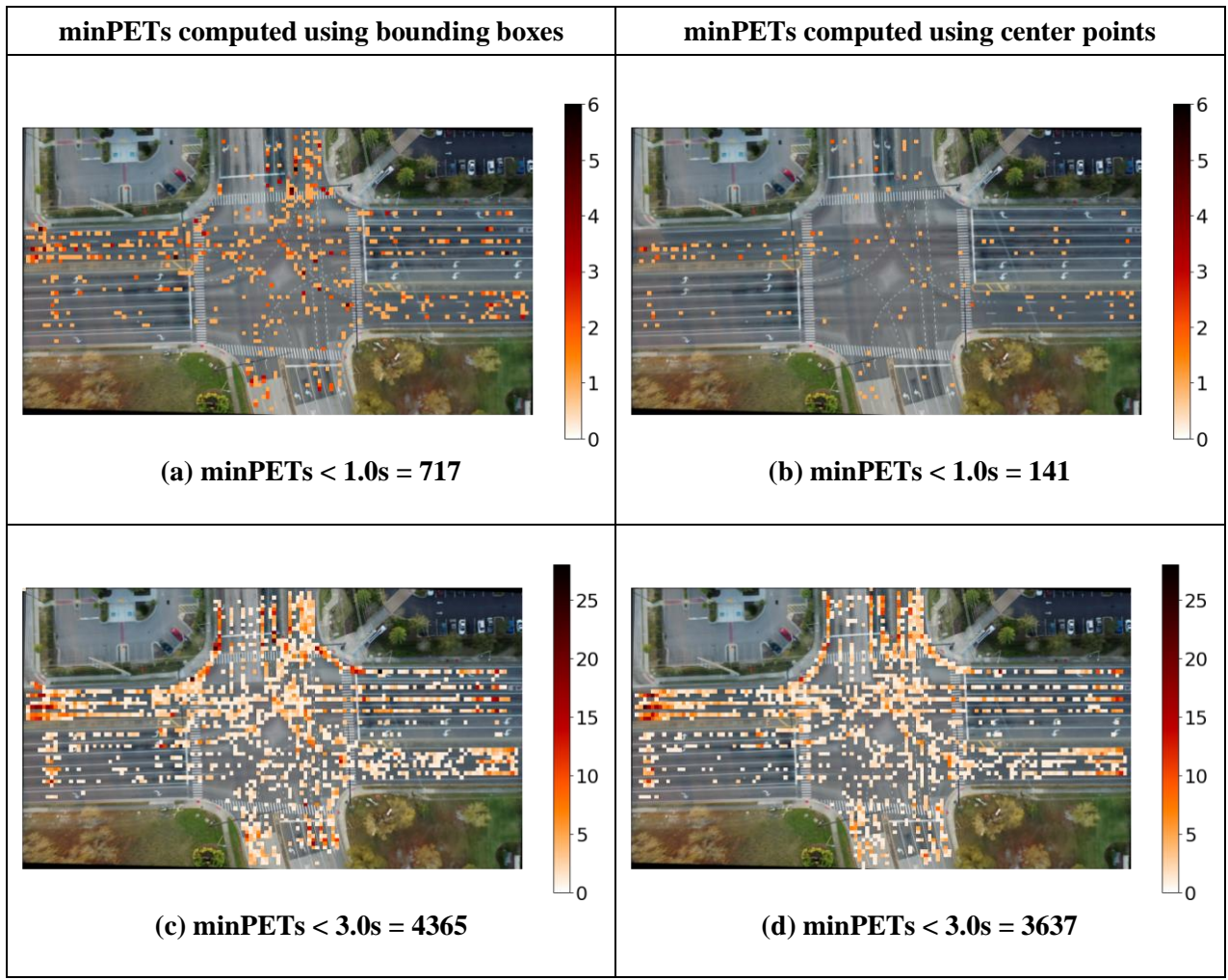

| **minPETs computed using bounding boxes** | **minPETs computed using center points** |
|---|---|
| **(a) minPETs < 1.0s = 717** | **(b) minPETs < 1.0s = 141** |
| **(c) minPETs < 3.0s = 4365** | **(d) minPETs < 3.0s = 3637** |

**Figure 3. minPET heatmaps computed using bounding boxes versus center points**

All the different datasets involving PET, speed, heading and signal timing was merged together to obtain the final dataset. The descriptive statistics of the different variables as well as brief explanation of each variable in the final dataset are shown in Table 2. The various signal timing such as red, green, yellow, etc are modelled as a countdown timer to understand the impact of the time remaining of a phase on PETs.

**TABLE 2 Variable Statistics**

| Feature | Description | count | mean | std | min | max |
|---|---|---|---|---|---|---|
| PET level | 1, if PET between 0.3s and 1s<br>2, if PET between 1s and 2s<br>3, if PET between 2s and 3s<br>4, if PET between 3s and 4s<br>5, if PET between 4s and 5s | 78859 | 3.3 | 1.18 | 1 | 5 |
| Distance (ft) | Spatial gap between two vehicles | 78859 | 13.98 | 1.53 | 0 | 15 |
| red_clearance (s) | Red clearance time remaining at the end of each phase | 78859 | 0.2 | 0.76 | 0 | 4.9 |
| all_red (s) | All red time remaining at the end of each cycle | 78859 | 0.02 | 0.22 | 0 | 3.9 |
| Red (s) | Red time remaining | 78859 | 6.11 | 25.28 | -1 | 175.9 |
| Yellow (s) | Yellow time remaining | 78859 | 0.01 | 0.61 | -1 | 4.9 |
| Green (s) | Green time remaining | 78859 | 19.25 | 16.8 | -1 | 85.3 |
| Phase 1 | | 78859 | 0.13 | 0.34 | 0 | 1 |
| Phase 2 | | 78859 | 0.32 | 0.46 | 0 | 1 |
| Phase 3 | | 78859 | 0.12 | 0.33 | 0 | 1 |
| Phase 4 | 1, if phase is active,<br>0, otherwise | 78859 | 0.02 | 0.15 | 0 | 1 |
| Phase 5 | | 78859 | 0.03 | 0.19 | 0 | 1 |
| Phase 6 | | 78859 | 0.09 | 0.29 | 0 | 1 |
| Phase 7 | | 78859 | 0.11 | 0.32 | 0 | 1 |
| Phase 8 | | 78859 | 0.06 | 0.24 | 0 | 1 |
| Speed (mph) | Current vehicle Speed | 78859 | 19.48 | 10.43 | 0 | 59.98 |
| Heading (degrees) | Direction of travel | 78859 | 190.81 | 100.83 | 0 | 360 |
| lane | Lane information for any PET | 78859 | 28.3 | 10.31 | 3 | 35 |
| Volume | Number of vehicles per 5 mins | 78859 | 49.1 | 11.68 | 19 | 86 |
| intersection | 1, if the vehicle is at intersection<br>0, otherwise | 78859 | 0.6 | 0.49 | 0 | 1 |
| speeding_prop | $\frac{speed-speed\ limit}{speed\ limit}$ for leading vehicle | 78859 | -0.45 | 0.24 | -0.89 | 0.33 |
| movement | Location of the vehicle<br>0, left turning lane<br>1, through lane<br>2, at intersection | 78859 | 1.48 | 0.7 | 0 | 2 |

A sample case of changing PETs towards the end of a cycle is shown in Figure 4. The PETs between interacting vehicles are shown in the figure. The lower the PET, the redder is the bounding box indicating high severity. It can be noted that as the phase turns green the vehicles start to move with PETs between 1.5 to 2s. As the phase turns from yellow to red, the PET even lowers to 0.8s as drivers try to clear the intersection.

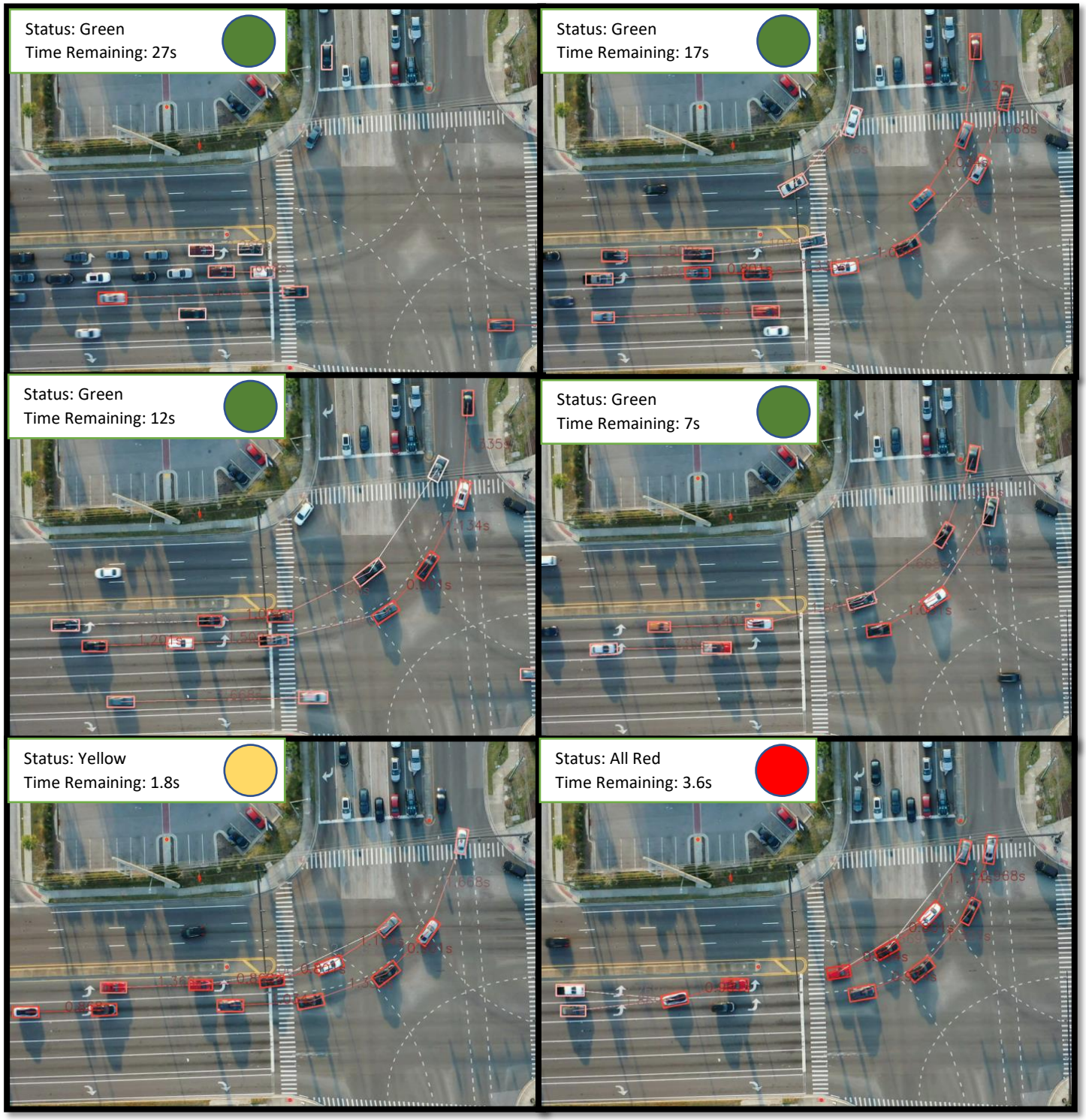

**Figure 4 Sequence of signal timing events showing decrease in PET for phase 1**

## MODEL
### Random Parameter Ordered Logit Model

Random parameters logit model is a logit model for which the parameters are assumed to vary from one individual to another. It is therefore a model that takes the heterogeneity of the population into account. In this study five levels of PET were considered.

We follow Milton et al. (2008) and Washington et al. (2011), and start with

$$U_{ij} = \beta_j X_{ij} + \varepsilon_{ij}$$

where $U_{ij}$ is a function determining the PET level i on individual PET for observaions, $X_{ij}$ is a vector of explanatory variables; βi is a vector of estimable parameters for outcome i which may vary across observations, and $\varepsilon_{ij}$ is the error term which is assumed to be generalized extreme value distributed (McFadden, 1981).

In order to develop random parameter models, we consider the following latent process as described by Sarrias Mauricio, 2016

$$y_{it}^* = x_{it}^T \beta_i + \epsilon_{it}, i = 1, \cdots, n; \ t = 1, \cdots T_i$$
$$\beta_i \sim g(\beta_i | \theta |), \tag{1}$$

where $y_{it}^*$ is a latent (unobserved) process for individual i in period t, $x_{it}$ is a vector of covariates, and $\epsilon_{it}$ is the error term.

Note that the conditional probability density function (PDF) of the latent process $f(y_{it}^* | x_{it}, \beta_i |)$ is determined once the nature of the observed $y_{it}$ and the population PDF of $\epsilon_{it}$ is known. If $y_{it}$ is binary and $\epsilon_{it}$ is distributed as normal, then the latent process becomes the traditional probit model; if $y_{it}$ is an ordered categorical variable and $\epsilon_{it}$ is logistically distributed, then the traditional ordered logit model arises. Formally, the PDF for binary, ordered, and Poisson model are, respectively

$$
\begin{aligned}
f(y_{\{it\}}^* | x_{\{it\}}, \beta_i |) &= \left[F(x_{\{it\}}^T \beta_i)\right]^{y_{\{it}} \left[1 - F(x_{\{it\}}^T \beta_i)\right]^{1-y_{it}} \\
&= \prod_{j=1}^{J} \left[F(k_j - x_{it}^T \beta_i) - F(k_{j-1} - x_{it}^T \beta_i)\right]^{y_{itj}} \\
&= \frac{1}{y_{it}!} \exp\left[-\exp(x_{it}^T \beta_i)\right] \exp(x_{it}^T \beta_i)^{y_{it}}
\end{aligned}
\tag{2}
$$

For the binary and ordered models, F (·) represents the cumulative distribution function (CDF) of the error term, which $F(\epsilon) = \Phi(\epsilon)$ for probit and $F(\epsilon) = \Gamma(\epsilon)$ for logit. For the ordered model, $\kappa_j$ represents the threshold for alternative j = 1, . . ., J−1, such that $\kappa_0 = -\infty$ and $\kappa_0 = \infty$.

In the structural model given by Equation 1, we allow the vector coefficient $\beta_i$ to be different for each individual in the population. In other words, the marginal effect on the latent dependent variable is individual-specific. Nevertheless, we do not know how these parameters vary across observations. All we know is that they vary according to the population PDF $g(\beta_i | \theta |)$ where $\theta$ represents the moments of the distribution such as the mean and the variance, which must be estimated. A fully parametric model arises once $g(\beta_i | \theta |)$ and the distribution of $\epsilon$ are specified.

For simplicity in notation, assume that the coefficient vector is independent normal distributed, so that $\beta_k \sim N(\beta_k, \sigma_k^2)$ for the k-th element in i. Note that each coefficient can be written as $\beta_{ki} = \beta_k + \sigma_k \omega_i$ where $w_i \sim N(0, 1)$, or in vector form as $\beta_i = \beta + L \omega_i$, where L is a diagonal matrix that contains the standard deviation parameters, $\sigma_k$. All the information about the individual heterogeneity for each individual attribute is captured by the standard deviation parameter $\sigma_k$. If $\sigma_k = 0$, then the model is reduced to the fixed parameter model, but if it is indeed significant then it would reveal that the relationship between $x_{itk}$ and $y_{it}$ is heterogeneous and focusing just on the central tendency k alone would veil useful information. It is useful to note that the random effect model is a special case in which only the constant is random.

Some measures of goodness-of-fit including Log-Likelihood and Akaike Information Criterion (AIC) were used to find the best fitted model. The best fitted model that displayed the maximum value of the log-likelihood function was chosen to obtain the parameter estimates that made the data most likely. AIC value was used to compare the performances of the GLMs. The preferred model is the one with the minimum AIC value. The AIC value can be evaluated using:

$$AIC = 2k - 2\ln(L)$$

where, k = The number of estimated parameters in the model, L = The maximum value of the likelihood function for the model.

The Bayesian Information Criterion (BIC) values were also calculated to conclude the best model that describes the relationship between each crash type and the explanatory variables. The AIC introduces a penalty term that is represented by the parameter number in the AIC. The BIC introduces the penalty term as a combination between the parameter number and sample size (26).

**RESULTS**

The results from the Random Parameter Logit Model are shown in Table 3 and the conclusions are presented in the subsections. It is important to note here that the signal times are modelled as a countdown timer. For example, a yellow time of 4s means that the signal state is currently yellow and has 4s remaining. The reason the authors decided to model the signal timings as a countdown timer is that most of the times the drivers would try to speed up or slow down to comply with the end of the signal timings.

**Table 3 Models for PET levels on Signal timing**

| Yellow time | | | | | | |
|---|---|---|---|---|---|---|
| No. of Obs = 2955, Log Likelihood = -3667, AIC = 7361, BIC = 7445 | | | | | | |
| Coefficients: | Estimate | Odds ratio | Std. error | z-value | Pr(>\|z\|) | |
| constant | 4.685 | - | 0.884 | 5.301 | 0.000 | *** |
| distance | -0.175 | 0.839 | 0.048 | -3.661 | 0.000 | *** |
| yellow | 0.151 | 1.163 | 0.035 | 4.359 | 0.000 | *** |
| phase 3 | 0.728 | 2.071 | 0.321 | 2.269 | 0.023 | * |
| phase 5 | -0.897 | 0.408 | 0.474 | -1.893 | 0.058 | . |
| phase 6 | 2.004 | 7.416 | 0.550 | 3.641 | 0.000 | *** |
| phase 7 | 1.444 | 4.238 | 0.288 | 5.022 | 0.000 | *** |
| phase 8 | 2.032 | 7.628 | 1.005 | 2.022 | 0.043 | * |
| mean. volume | 0.010 | 1.010 | 0.013 | 0.745 | 0.456 | |
| std. dev volume | 0.056 | 1.058 | 0.003 | 18.970 | 0.000 | *** |
| kappa.1 | 3.853 | - | 0.117 | 32.892 | 0.000 | *** |
| kappa.2 | 5.298 | - | 0.130 | 40.651 | 0.000 | *** |
| kappa.3 | 7.004 | - | 0.150 | 46.577 | 0.000 | *** |

| | | | | | |
|---|---|---|---|---|---|
| **All Red time** | | | | | |
| No. of Obs = 805, Log Likelihood = -683.8, AIC = 8441, BIC = 8567 | | | | | |
| **constant** | 14.475 | - | 1.162 | 12.451 | 0.000 | *** |
| **Speeding prop** | 3.956 | 52.274 | 0.732 | 5.404 | 0.000 | *** |
| **All red** | -0.558 | 0.572 | 0.124 | -4.515 | 0.000 | *** |
| **phase 1** | -2.573 | 0.076 | 0.570 | -4.514 | 0.000 | *** |
| **phase 2** | 3.644 | 38.238 | 0.605 | 6.027 | 0.000 | *** |
| **phase 3** | -1.262 | 0.283 | 0.573 | -2.202 | 0.028 | * |
| **phase 4** | 4.986 | 146.288 | 1.332 | 3.743 | 0.000 | *** |
| **phase 5** | 4.196 | 66.430 | 1.122 | 3.739 | 0.000 | *** |
| **phase 7** | 4.058 | 57.855 | 0.861 | 4.711 | 0.000 | *** |
| **mean. volume** | -0.089 | 0.915 | 0.018 | -5.022 | 0.000 | *** |
| **std. dev volume** | 0.117 | 1.124 | 0.007 | 15.714 | 0.000 | *** |
| **kappa.1** | 4.960 | - | 0.438 | 11.316 | 0.000 | *** |
| **kappa.2** | 8.259 | - | 0.515 | 16.039 | 0.000 | *** |
| **kappa.3** | 11.026 | - | 0.596 | 18.487 | 0.000 | *** |
| **Red clearance** | | | | | | |
| No. of Obs = 6785, Log Likelihood = -8881, AIC = 19792, BIC = 19915 | | | | | |
| **constant** | 8.072 | - | 0.574 | 14.058 | 0.000 | *** |
| **Speeding prop** | 0.421 | 1.523 | 0.159 | 2.654 | 0.008 | ** |
| **distance** | -0.155 | 0.856 | 0.023 | -6.672 | 0.000 | *** |
| **Red clearance** | 0.033 | 1.033 | 0.023 | 1.443 | 0.149 | |
| **intersection** | -0.338 | 0.714 | 0.084 | -4.003 | 0.000 | *** |
| **phase 1** | -0.701 | 0.496 | 0.413 | -1.696 | 0.090 | . |
| **phase 2** | -1.030 | 0.357 | 0.209 | -4.937 | 0.000 | *** |
| **phase 3** | -1.821 | 0.162 | 0.605 | -3.010 | 0.003 | ** |
| **phase 4** | -5.175 | 0.006 | 0.703 | -7.360 | 0.000 | *** |
| **phase 5** | -3.126 | 0.044 | 0.329 | -9.492 | 0.000 | *** |
| **phase 6** | -1.471 | 0.230 | 0.230 | -6.402 | 0.000 | *** |
| **phase 7** | -1.488 | 0.226 | 0.218 | -6.820 | 0.000 | *** |
| **phase 8** | -1.106 | 0.331 | 0.319 | -3.466 | 0.001 | *** |
| **mean. volume** | -0.001 | 0.999 | 0.010 | -0.063 | 0.950 | |
| **std. dev volume** | 0.042 | 1.043 | 0.002 | 25.986 | 0.000 | *** |
| **kappa.1** | 3.514 | - | 0.083 | 42.558 | 0.000 | *** |
| **kappa.2** | 4.850 | - | 0.089 | 54.498 | 0.000 | *** |
| **kappa.3** | 6.514 | - | 0.098 | 66.606 | 0.000 | *** |
| **Red Time** | | | | | | |
| No. of Obs = 7112, Log Likelihood = -10150, AIC = 20334, BIC = 20451 | | | | | |
| **constant** | 4.198 | - | 0.222 | 18.950 | 0.000 | *** |
| **Speeding prop** | 1.398 | 4.049 | 0.105 | 13.264 | 0.000 | *** |
| **distance** | 0.096 | 1.101 | 0.011 | 9.036 | 0.000 | *** |
| **red** | -0.004 | 0.996 | 0.001 | -7.579 | 0.000 | *** |
| **intersection** | -1.043 | 0.352 | 0.070 | -14.912 | 0.000 | *** |
| **phase 2** | 0.371 | 1.450 | 0.089 | 4.158 | 0.000 | *** |
| **phase 4** | 0.639 | 1.894 | 0.163 | 3.921 | 0.000 | *** |
| **phase 6** | 0.652 | 1.920 | 0.134 | 4.868 | 0.000 | *** |
| **mean. volume** | -0.011 | 0.989 | 0.002 | -4.674 | 0.000 | *** |
| **std. dev volume** | 0.013 | 1.013 | 0.003 | 4.616 | 0.000 | *** |
| **kappa.1** | 2.400 | - | 0.084 | 28.496 | 0.000 | *** |
| **kappa.2** | 3.659 | - | 0.122 | 30.101 | 0.000 | *** |
| **kappa.3** | 4.859 | - | 0.157 | 30.943 | 0.000 | *** |
| **Green Time** | | | | | | |
| No. of Obs = 62416, Log Likelihood = -91730, AIC = 183497, BIC = 183651 | | | | | | |
| **constant** | 4.767 | - | 0.092 | 51.701 | 0.000 | *** |
| **Speeding prop** | 0.850 | 2.340 | 0.033 | 25.571 | 0.000 | *** |
| **distance** | -0.082 | 0.921 | 0.005 | -14.983 | 0.000 | *** |
| **green** | 0.019 | 1.020 | 0.001 | 28.633 | 0.000 | *** |
| **intersection** | -0.272 | 0.762 | 0.016 | -16.767 | 0.000 | *** |
| **phase 2** | -0.265 | 0.767 | 0.027 | -9.951 | 0.000 | *** |
| **phase 4** | -0.922 | 0.398 | 0.050 | -18.607 | 0.000 | *** |
| **phase 5** | 0.146 | 1.157 | 0.046 | 3.202 | 0.001 | ** |
| **phase 6** | -0.641 | 0.527 | 0.035 | -18.238 | 0.000 | *** |
| **phase 7** | 0.174 | 1.190 | 0.032 | 5.374 | 0.000 | *** |
| **phase 8** | -0.606 | 0.546 | 0.039 | -15.690 | 0.000 | *** |
| **mean. volume** | -0.001 | 0.999 | 0.001 | -1.206 | 0.228 | |
| **std. dev volume** | 0.006 | 1.006 | 0.001 | 4.394 | 0.000 | *** |
| **kappa.1** | 2.511 | - | 0.024 | 103.157 | 0.000 | *** |
| **kappa.2** | 3.531 | - | 0.032 | 110.856 | 0.000 | *** |
| **kappa.3** | 4.712 | - | 0.041 | 115.072 | 0.000 | *** |

It was seen that the random parameters models performed better than the fixed effects models as the AIC and BIC values of the random parameter models were much lower than those of the fixed effects models. The study evaluated the effect of signal times on PET levels. Thus, five models for different signal times (yellow, all red, red clearance, red and green) were performed to ascertain its effects on PETs. As mentioned before, PETs less than 1 was indicated as level 1, and the data had five levels of PET, with PET values ranging from 0.3secs to 4.97 secs. Other independent variables in the models were the different phases for the signal cycle, phase 1 through 8, where phases 2,4,6,8 were for through and right turn and phases 1,3,5,7 were left turning ones. The negative signs of the coefficients indicate that the presence or increase in that variable has potential to reduce the PET level.

**Intersection vs PET**

PET values also have different effects when vehicles are inside the intersection verses when they were in the approach. It can be seen that for the models of red clearance, red and green timings, the intersection indicator variable was significant and the coefficient being negative indicates that at intersections the PET levels are in general low meaning the vehicles have tendency to maintain low gaps between them, which can in turn be a risky situation.

**Yellow Time vs PET**

The overall yellow time is positively related to the different PET. The lower the yellow time, the lower the PET which shows that the vehicles tend to follow each other closely towards the end of the yellow phase. The variable phase 5 shows that when the yellow for this phase is active, there are low PETs. This essentially indicates a probable issue with the length of the yellow time. The other phases that came out to be significant have the opposite relationship and can be interpreted to be safer.

**All Red Time vs PET**

All red time is negatively correlated to PET. This shows that the vehicles that enter the intersection at the end of yellow have lower PET since they are essentially trying to clear the intersection. Together with the yellow time and all red time, it can be concluded that there are lower PETs at the boundary of yellow and all red time. The variables phase 1 and phase 3 have negative sign meaning that when the all red of these phases are active, there are lower PETs resulting in an unsafe state. This also helps to conclude the visualization in Figure 2, where we see a snapshot of the traffic state for phase 1 at all red time of 1.8s.

**Red Clearance Time vs PET**

The red clearance time was not significant in the model but from the individual red clearance time per phase it is noted that the relationship is negative meaning that each of the clearance times experience lower PET. This can also be noted as a potential safety condition that will require careful signal timing optimization. Almost all phases except phase 1, were found to be significant.

**Red Time, Green Time vs PET**

It can be seen that increase green signal times of a cycle have positive signs indicating potential for increasing PET level. Which in turn signifies that increase in these timings have potential to increase PET values between vehicles and increasing safety by reducing probability of conflict leading to rear end crashes. On the other hand, increase in red times, influences the PET levels to decreases, meaning that the increase in these timings have potential to decrease the PET values between vehicles as drivers become impatient and try to choose unsafe gaps which increases the probability of rear end crashes.

**Speeding Proportion vs PET**

Increase of speed of the vehicle also leads to an increase in PET. The speeding proportion is calculated for the leading vehicle and as such once this vehicle speeds, the distance between interacting vehicles increases thus increasing PET.

**Odds Ratio**

Standard interpretation of the ordered logit coefficient is that for a one unit increase in the predictor, the response variable level is expected to change by its respective regression coefficient in the ordered log-odds scale while the other variables in the model are held constant. Thus, we calculate Odds ratio. Odds Ratios can be obtained by exponentiating the ordered logit coefficients, $e^{coef}$. For a one unit change in the predictor variable, the odds for cases in a group that is greater than k versus less than or equal to k are the proportional odds times larger, where k is the level of the response variable. So, as the coefficients of all red and red timings were negative, with one unit increase in all red and red (when other variables are constant), the odds of low PET meaning high risk values are 0.572 and 0.992 times larger respectively. So, for yellow, red clearance and green timings, as these coefficients are positive, with one unit increase in yellow, red clearance and green timings (when other variables are constant in each of the models), the odds of values in high PET meaning low risk levels are 1.163, 1.033 and 1.020 times larger respectively. This leads to an important conclusion regarding improving the PETs at intersections. Increasing the yellow, red clearance and green timings would lead to better PETs than increasing all red and red time. Since the data collected was during the afternoon peak, it might also be impactful to increase yellow and red clearance time for these periods only rather than for the entire length of day.

## CONCLUSION

In summary, this paper proposes the use of UAV vehicle trajectory data to understand the relationship between signal timing and PET. One hour of UAV data was collected to obtain PETs, speeding, heading and signal phasing and timing. The PETs were calculated using rotating bounding boxes and also using the back of the leading vehicle and front of the lagging vehicle which gives a much accurate PET than that using center points of the vehicles. It was then modelled using Random Parameter Ordered Logit Model. The PET values were divided into five classes. Results from the model showed yellow time and red clearance time is negatively related with PET while all red time, red time and green time are positively related. The odds ratio indicated that it would be possible to increase the PET levels and thereby improving the safety by only increasing the yellow time and red clearance time by 1 second.

This study can be used to understand the safety of an intersection in terms of signal timing. It will give insights as to whether the signal retiming will help improve safety. Only an hour of video data processing has the potential to provide these insights to relevant authorities. Future studies can focus on the traffic dynamic features as well as different types of intersections to understand the relationship between surrogate safety measures and signal timing.

## AUTHOR CONTRIBUTIONS

The authors confirm contribution to the paper as follows: study conception and design: Zubayer Islam, Mohamed Abdel-Aty, Amrita Goswamy, Amr Abdelraouf; data collection: Ou Zheng, Amr Abdelraouf, Zubayer Islam; analysis and interpretation of results: Zubayer Islam, Mohamed Abdel-Aty, Amrita Goswamy. Author; draft manuscript preparation: Zubayer Islam, Mohamed Abdel-Aty, Amrita Goswamy, Amr Abdelraouf. All authors reviewed the results and approved the final version of the manuscript.